How shall we use the proteomics toolbox for biomarker discovery?

Pierre Lescuyer [1,2], Denis Hochstrasser [1,2], Thierry Rabilloud [3,4]*

1. Department of Structural Biology & Bioinformatics & Pharmacy Section
Geneva University Faculties of Medicine & Sciences

2. Laboratory Medicine Service
Department of Genetic & Laboratory Medicine
Geneva University Hospital
CH-1211 Geneva 14, Switzerland

3. CEA-DSV-iRTSV/BBSI
CEA Grenoble
17 rue des martyrs
F-38054 Grenoble Cedex 9 France

4. CNRS UMR 5092
17 rue des martyrs
F-38054 Grenoble Cedex 9 France

* to whom correspondence should be addressed
Thierry  Rabilloud
iRTSV/BBSI, UMR CNRS 5092
CEA Grenoble
17 rue des martyrs
F-38054 Grenoble Cedex 9 France
Mail: Thierry Rabilloud@cea.fr
________________________________

Abstract

Biomarker discovery for clinical purposes is one of the major areas in which proteomics is used. However, despite considerable effort, the successes have been relatively scarce. In this perspective paper, we try to figure out and analyze the main causes for this limited success, and to suggest alternate strategies, which will avoid them, without eluding the foreseeable weak points of these strategies. Two major strategies are analyzed, namely the switch from body fluids to cell and tissues for the initial biomarker discovery step or, if body fluids must be analyzed, the implementation of highly selective protein selection strategies.

Synopsis : this paper aims at stimulating discussions on how proteomics could be best used for biomarker discovery. From the sucesses and failures described in the literature, current approaches are critically discussed, and strategies for a better use of proteomics in biomarker discovery are outlined.

___________________________________________________________________________
_

1. Frame and Facts.

The present paper does not aim at being a review on biomarkers, as this has been already done (e.g. in Hu et al. [1]). It is an opinion paper aiming at provocative discussion on how proteomics is more or less suited to the discovery of new biomarkers, and how the power of proteomics can be used in improved strategies to maximize the probability of new markers discovery. This restricted frame implies in turn that the discussion on what is a biomarker is outside the scope of this paper. As a matter of facts, this issue has also been reviewed recently in the frame of proteomics studies [2]

Because of their easy accessibility compared to tissue biopsies, biological fluids are targets of choice for looking for diagnostic, prognostic and even treatment follow up biomarkers. The paradigm is that the tissues, which are in contact with biological fluids, are prone to liberate protein components in the fluids, and that the disease-altered state will change either the spectrum or the amount of liberated proteins. Ideally, such proteins can be disease-associated biomarkers and, consequently, proteomic analysis of body fluids should reveal a lot of new diagnostic markers. However, there is obviously a huge gap between theory and reality. This statement is clearly illustrated when comparing the enormous amount of data generated until now by proteomics studies focusing on body fluids and the virtual absence of concrete results in term of discovery of new clinically applicable biomarkers. How can we explain this failure? The reason is probably not that the paradigm is false. Indeed, we know that a class of biomarkers of special interest is constituted of tissue leakage proteins, i.e. proteins that are liberated in fluids by diseased tissue differently than healthy tissue. This class of biomarkers traces back in the past, e.g. with the liver transaminases which increased level signs destruction of liver cells [3], but more recent examples are available, e.g. the prostate specific antigen (PSA) for prostate cancer [4] or the troponin I and T for acute myocardial infarction [5,6]. However, such markers exist in limited numbers and in quite low concentration, and it can be hoped that a wide search for tissue leakage markers should provide some new and interesting candidates. Most likely, the path to success will rely on the separation of two discovery and validation phases.

2. Success & Failure.
Accordingly, a few success stories can be attributed to proteomics in this area, one of the earliest being the discovery of 14-3-3 proteins in cerebrospinal fluid (CSF) as markers of certain types of brain destruction, including Creutzfeldt-Jakob disease (CJD) [7]. Although this marker has been challenged for its selectivity and specificity, it remains one of the rare successes of proteomics in the field of biomarkers discovery. However, it should be reminded first that CSF contains much less proteins than plasma, 1/200, so that the "buffering" of protein composition is much weaker in CSF. Second, CSF collects protein mostly from one organ, the central nervous system and the spinal cord, and third, the mass ratio between this organ and the fluid volume is much more favorable in CSF than in plasma. Altogether, a brain leakage is expected to liberate higher amounts of proteins in a lesser volume, in a relatively simpler fluid, and this can make the difference between failure and success. Another recent example is the discovery of stroke biomarkers in the spinal fluid in a first discovery phase and their validation in serum in a second clinical validation phase. The situation is much bleaker for plasma-derived biomarker discovery. Very important efforts have been devoted to find cancer-associated biomarkers, for results of limited interest. Most studies involving the comparison of plasmas obtained from healthy donors with plasmas obtained from cancer-diagnosed donors have shown essentially inflammation-related proteins (e.g. haptoglobins or serum amyloid protein) as differential markers [8-12], which shows in turn that these markers are indeed not disease-specific. It then appears clearly that among the various biological fluids, we can oppose different types having potency and use widely different from one to another.

3. Different biological fluids.
Fluids collecting proteins from one particular organ or anatomic structure, such as CSF, articular fluid or pleural fluid, despite their relatively very high complexity similar to plasma, results in an increased probability of finding biomarkers related to the associated organ because of the organ proximity and a smaller dilution effect. The problem is that these fluids have to be collected in a somewhat invasive manner, not so different from a needle tissue biopsy in fact. However, in the case of CSF at least, the corresponding organ (brain) is otherwise most difficult to analyze by biopsies, especially on healthy control subjects. In this sense, these fluids are of interest although for only a limited spectrum of diseases. In addition, their rather limited availability disqualifies them as a source for early biomarker detection in a wide population of people with otherwise limited symptoms. A second category of body fluids is represented by excretory mediums, urine or bile. Urine especially is rather easy to collect and available in large amount, what could be of interest for proteomic applications. The problem is that, by definition, these fluids do not have a homogenous composition. Protein concentration in urine can change from milligrams to grams per liter depending on the disease state. In addition, these fluids contained various (usually high) concentrations of excreted compounds, such as salts, urea or solubilizing agents. Both aspects make them very difficult to analyze by comparative proteomics. Another difficulty with urine comes from the fact that, in addition to proteins collected from a few organs involved in its production and excretion (kidney and urinary tract, including bladder), a large part of the proteome originates from plasma (through glomerular filtration or leakage and tubular secretion or non reabsorption). The third type of body fluids corresponds to plasma/serum, which was presented as the most amenable to the disease-associated biomarker discovery process. This fluid is, indeed, easily collected, and the important point is that plasma collects proteins from each and every tissue, and should therefore be a "universal" source of biomarkers, opposite to other fluids such as urine and CSF. As a matter of facts, most protein biomarkers used to date in clinical practice are measured in plasma, and most of this paper will be devoted to this source of biomarkers.

4. Proteomic strategy.
However, the conclusion that has to be driven at that point is that the current proteomics strategy used for biomarker discovery, i.e. comparing plasma protein profiles in normal versus diseased subjects, has proven its inefficiency or even failure during the discovery phase. There are many reasons for that, but one of the most obvious is the tremendous dynamic range of protein levels in plasma, ranging from below the nanogram per ml to tens of milligrams per ml, which is twelve orders of magnitude (from millimolar concentration of albumin to fentomolar concentration of tumor necrosis factor TNF). Indeed, 22 plasma proteins make 99% of the protein mass. It is therefore tempting to try to deplete these proteins to enhance the representation of minor proteins and thus to bring them above the threshold of detection by proteomics. Although it has received considerable attention and development, this approach has also failed, for a range of reasons. First of all, because of this enormous dynamic range, the removal process must be close to 100% efficient to be interesting. Even though, a perfect process removing the 22 abundant proteins (which is far from being achieved) will leave 8 to 10 orders of magnitude of protein levels in plasma, which also exceeds by far the current power of proteomics technologies that covers 4 orders of magnitude. In addition, less abundant interesting proteins may also be eliminated during the depletion process, resulting in loss of reliable information and potential artefactual discrepancies between samples. In other words, it is our opinion that this process is bound to fail in the discovery phase. This opinion is not restricted to a peculiar setup in proteomics. While gel-based proteomics is mostly used and has failed in this respect, shotgun proteomics has not performed much better [13], and peptide selection by binding technologies such as SELDI has not found tissue leakage markers in plasma, although this has been claimed in CSF [14]. This is further exemplified by the fact that even the well characterized tissue leakage biomarkers have not been rediscovered by proteomics when the proteomics toolbox has been applied to the disease which they characterize (e.g. the liver transaminases in hepatitis) [15].

5. Definition & application of biomarker.
Beside these technical considerations, we believe that the current inability of body fluids proteomics to provide clinically useful markers further lies in a problem of definition: the expression "potential biomarker" does not have the same meaning for a proteomist, a clinical chemist or even a clinician at the bed side. The word biomarker itself is clearly defined [16, 17] and its meaning should be consensual. However, this is the notion of potentiality associated with the word biomarker that is matter to discussion. Typical proteomic studies on body fluids end up with a list of tens to hundreds differentially expressed proteins. For a proteomist, they represent potential biomarkers. For a clinical chemist, until proven otherwise, this is still a list of proteins. The question is: what makes the potentiality? In clinical practice, a biomarker has to be evaluated by a number of criteria, including sensitivity, specificity, positive and negative predictive values which rely upon the prevalence of the disease of interest. Determination of these parameters involves studies on large cohorts of hundreds or thousands of patients during the clinical validation phase. Furthermore, these criteria are valid in the context of a well-defined pathological situation. The quantitative method (usually an immunoassay) used for such studies itself needs to fit to precise criteria in term of precision, accuracy or linear reportable range. This means that the proteomic study is just the beginning of the story and that proteomists should provide more than lists of differentially expressed proteins if they want to convince clinical chemists and physicians of the usefulness of their science for discovery of new biomarkers. This requirement for proteomic workflows to fulfill clinical studies criteria in order to focus the biomarker selection process on the most valuable candidates is highlighted in a recent paper [2]. The problem is that validation studies represent huge amount of work for a single protein. So what about, for tens or hundreds of proteins? The story of 14-3-3 proteins illustrates well the length and the complexity of the way between proteomic discovery and application into clinics. In the original study [18], hundreds of gels were ran to show that two differentially expressed spots, latter identified as 14-3-3 proteins [7], were potential biomarkers of CJD. Then, numerous studies using specific immunoassays and involving hundreds of patients were performed to challenge the clinical utility of the 14-3-3 proteins. Interestingly, the conclusion was that despite good sensitivity (90-97%) and specificity (87-100%) [19], the detection of 14-3-3 proteins in CSF is of limited clinical interest, notably because, due to the low prevalence of CJD, it has a poor positive predictive value [20, 21].

Then, what are the characteristics of a good biomarker? First, it should be as specific as possible for a given disease. In other words, it should have a low false positive rate. For example, many diseases have an inflammatory dimension, and it is therefore not very surprising to find a difference in the inflammation-related markers in a comparison between healthy subjects and subjects suffering from a variety of diseases. Recent studies have also described fragments of classical plasma proteins as cancer-related biomarkers, and it has been advocated that specific cleavage of these proteins by the proteases released by cancer cells to invade the surrounding tissues may explain these observations [22]. However, such proteolysis-derived markers pose a number of problems. For example, they will be difficult to diagnose by standard, antibody-based methods. In addition, their robustness is expected to be limited. As a matter of facts, matrix metallo-proteases (MMPs) have been put forward as the cancer-specific proteases responsible for these specific protein fragments. However, it should be kept in mind that some cell types (e.g. macrophages, also produce MMPs, although in lesser amounts. Thus, aging of a sample rich in macrophage-derived products (e.g. by inflammation) is very likely to produce such fragment markers.

Second, and this is frequently overlooked, the biomarker should be robust. The clinical practice is in this sense a bit remote from the situation prevailing in research laboratories, and some markers, which appear relevant at the research laboratory level in very controlled sample collection and processing conditions are not amenable to testing at the bedside, where clinical practice constraints bring some variability, especially with sample processing and thus sample degradation.

Then, for each disease, the marker should show a marked difference between the healthy, another disease and the diseased populations of interest, so that the disease can be specifically and selectively diagnosed from other pathological conditions.

Finally, the marker should be useful for clinical usage, i.e. afford an interesting service for the clinicians compared to other investigation methods. For example, in the field of breast cancer, a biomarker which would be not diagnostic for tumors of a size smaller that the one detected by mammography is of limited interest, unless it offers a cost advantage or another added value, e.g. a prognostic value or a population survey applicability.

6. What to do?

This does not mean at all, however, that proteomics-based biomarker research should stop. It just means that it is useless to make it the way it has been made up to now in many laboratories and that it should be made differently. In particular, proteomic studies have to be designed to maximize the likelihood of identifying proteins having a real potential as disease-associated biomarkers and, thereby, facilitate the selection process for further costly and time-consuming clinical studies. Each proteomic study aiming at the discovery of biomarkers should then include two distinct and complementary steps: a discovery phase and a validation phase. The discovery phase corresponds to the classical proteomic analysis, involving protein fractionation, identification and quantitation using gel- and/or liquid chromatography- and MS-MS-based technologies. Concerning this primary phase, the current inefficiency of proteomics-based biomarker research indicates that we will probably have to change drastically the way we look for biomarkers. Accordingly, we would like to propose several simple rules.

6.1. A relevant question.

Firstly, the clinical question motivating the research should be carefully defined. This implies tight collaboration with clinical chemists and physicians to determine where there is a need for new biomarkers in term of diagnosis, prognosis, prevention, prediction, therapeutic selection or follow-up. A clear definition of the project and of the clinical benefit expected will allow choosing appropriate samples both for proteomic analysis in the discovery phase and for preliminary validation experiments in the clinical phase. It will also provide a clear frame for data interpretation and thus help selecting valuable candidate biomarkers. Collaboration with physicians is also mandatory to define proper sample collection procedures that are compatible with current clinical practices. In particular, problems of adherence of the medical and technical staff to the designed protocols should not be underestimated. For example, one requesting tissue samples from surgical resection should be aware of a possible delay in sample freezing within the operating room. It is crucial to control and monitor these pre-analytical variables since they can have a dramatic effect on quality and the reproducibility of the results [23][24].

6.2. Appropriate samples in the discovery phase.
Secondly, the discovery phase should be conducted on appropriate samples. Post mortem samples can even be used in this early phase. Post mortem samples have the advantage of being collectable on organs that are not available to sample collection of a living individual (e.g. brain). Moreover, the important proteolysis taking place after death maximizes the effects observed in sample collection procedures on living individuals, thereby selecting for the most robust markers (i.e. those surviving a strong proteolysis event).
In the brute force way that has been used in several situations up to now, many are directly trying to find interesting differences in the fluid without prior investigation of the tissue of interest. In the example of solid tumors, some are looking at the differences in the plasma without any serious characterization of the tumor tissue compared to the normal one. However, reversal of this paradigm is likely to lead to interesting results. By paradigm reversal, it is meant that one should first compare normal tissue to diseased tissue or even better diseased tissue with two or several well separated etiologies, find differences, and then look for these differences by focused approaches (mass spectrometry – MRM or antibody - immunoassay based) in biological fluids. Otherwise, one can also look for organ specific proteins using for example the human atlas of proteins, the human peptide atlas and SwissProt-UniProt, and once they are found, try to look for them in clinical situations in which specific organ destruction is expected. This latter model is just trying to reproduce what takes place for now well-accepted biomarkers (e.g. troponin) at the proteomics scale, and traces back to the protohistory of proteomics [25]. The comparison between differently diseased versions of the same tissue or cell type has received brilliant recent demonstration on the search for a metastatic power associated biomarker [26]. It can be argued that such a comparison has been carried out many times, with limited success. However, it must be kept in mind that most of the times, this comparison is intended to provide mechanistic clues for the disease, so that many proteins are excluded from the scope of interest. Furthermore, even though housekeeping proteins plague the analysis of cells in exactly the same way than major proteins in plasma, it should be kept in mind that the protein level dynamics in 6 to 8 orders of magnitude in cells, compared to the 10 or 12 orders of magnitude in plasma and 8 to 10 orders of magnitude in perfectly depleted plasma. Furthermore, the fact that few intracellular proteins are glycosylated enhances the separating power of protein separation in the case of cells versus plasma. All in all, this means that the technical progresses in proteomics will give us access to disease-typing proteins in cells well before the same level of performance will be reached in plasma. The price to pay for that is to replace the lack of power in protein analysis in proteomics by more clever and focused cell biology upfront, i.e. more use of animal and cellular models, and to take the bet for transposition to human disease. Even more, we can make the best possible use of the cellular or animal systems, for example by searching for differential proteins not in the complete cells or tissue, but in what is released in a suitable medium by the adequate cells or tissue explants. This is the strategy used for the so-called secretome studies. The fact that these conditioned media do contain a fair proportion of cytosolic proteins released by cell breakage is clearly a drawback when bona fide secreted proteins are to be studied [27,28]. However, it could become an advantage in a more general search for biomarkers [29]. For example, it has been shown recently that proteins present in the secretion medium at concentrations down to the low ng/ml range can be analyzed with standard proteomics tools [30]. Furthermore, "secretomes" are probably better samples than complete tissue or cell extracts, in the sense that there is in "secretome" samples a bias in favor of easily-liberated proteins and against major tissue proteins present in large complexes (e.g. cytoskeletal proteins)

6.3. Exception or clever tricks.
Thirdly, since each rule has exception, proteomic analysis of body fluids could be of interest in some cases. This could include analysis of neighboring fluids of a particular organ or tissue (e.g. CSF for markers of brain damage [31,32]). When carried on plasma or urine, the proteomic analysis should involve a fractionation step aiming at the selection of a subclass of proteins to carry out the biomarker search. One recent example is the restriction to glycoproteins [33]. In this work, the search is carried out not on the complete glycoproteins, but on protein-derived glycopeptides. This is likely to be an important point for the success of the approach. As a matter of facts, glycosylation brings an enormous chemical diversity to proteins, and the separation of glycoforms spreads widely in the protein-separation space. Conversely, the glycopeptide-selection approach by hydrazide trapping and glycosidase release of the peptides will convolute all the glycoforms of the same peptide in a single signal, thereby decreasing the noise introduced by the major glycoproteins. The recent proof of concept [34] shows that this approach looks promising. However, it can be prone to opposite criticism. The first criticism is that the glycoforms by themselves can be biomarkers, and this information is lost by this procedure. The second and opposite criticism is that the major mass of proteins in a cell is not glycosylated, so that the most abundant tissue leakage markers are likely to escape this type of detection. As an example, neither the transaminases nor the troponin are glycosylated.
This example of glycosylation is probably an example of a more generally applicable strategy for the direct search of markers in plasma, which could be defined as positive selection strategies. In this case, the strategy is to select the subgroup of N-glycosylated proteins. This positive selection strategy is conceptually opposed to negative selection strategies (e.g. depletion of major serum proteins), which are bound to fail.
It can be forecast that other positive selection strategies could be of interest, e.g. those based on active site classes [35]. However, due to the tremendous complexity of plasma, efficient strategies will be limited to those showing very high selectivity on the chemical sense of the word. Broad selection on general parameters, e.g. by hydrophobicity or ion exchange, has been applied numerous times (e.g. by the SELDI approaches) and has demonstrated its poor value. It can be also forecast that the combination of target tissue selection and of chemical selection of protein subgroups can bring added value in the search for biomarkers [36], at the expense of the broadness of the approach.
A special case in positive protein selective is represented by the equalizing beads approach [37,]

This combinatorial approach is for sure able to reveal proteins present at very low levels, and this is a highly attractive feature in biomarker discovery. However, the very principle of this approach is to provide a limited binding capacity for each and every protein on one or a few bead(s) present in the combinatorial process. This implies in turn that one the binding sites present have been saturated (and this should be reached at rather low amounts) any quantitative dimension is lost. This is a major drawback in biomarker discovery, as "normal" values for biomarkers are generally not equal to zero. Thus, it is our opinion that this approach will release its full potency in biomarker discovery when a clever trick for restoring the quantitative dimension will be found. Finally, careful bibliographic and databases searching has to be performed to find information that could bring added value, such as tissue specificity or a link with a physiopathological mechanism and thus help selecting for further validation studies. Here, we would like to insist on the fact that the wish of having bioinformatics tools capable to automatically search, integrate and interpret information obtained even by others investigators in related pathologies or on the same tissue is far from being granted, if it will ever be, The lack of standardization between experimental workflows, the great diversity of data resources, the irregular quality of data provided by databases explain, at least partly, this matter of facts. So, time, patience, perseverance and critical mind are still, and probably for a while, the main required qualities for reliable information searching. On the other hand, bioinformatics tools are obviously requested to help researchers in this difficult task: for data storage and management, for database searching, for statistical analysis, for integration of clinical, experimental and external data, and to help in their interpretation. Standardization of procedures and tools in this area should probably represent a way to promote and facilitate data exchange and comparison. However, this subject is out of the scope of this article.

6.4. Samples for the validation phase.
Once the discovery phase has been performed, it must be followed by a validation phase, which represents a full part of the biomarker identification workflow. The goal is to select from the list of proteins the ones with the highest potential considering the clinical problem that need to be solved. The sample of choice for such validation studies is obviously plasma/serum since this is the main sample in used in current clinical chemistry practice. Validation experiments, based on quantitative measurements, are carried out to confirm that the concentration of the candidate protein is significantly different between diseased and control state. They will provide preliminary values of sensitivity and specificity to determine whether the potential biomarker is amenable for large cohorts studies. Recently, MS-based techniques have been described allowing quantitative measurements of specific peptides that could be of interest for preliminary validations steps [38]. However, the gold standard techniques for such quantitative measurements are immunoassays, mainly ELISA, thank to the unmatched sensitivity and specificity provided by antibodies. As previously stated for the discovery phase, samples have to be carefully chosen to answer to a precise clinical question. In particular, control samples should include pathologies related to the disease investigated. For example, in the context of CJD diagnosis, control samples should include other type of neurodegenerative dementias. It would be also of interest to investigate susceptibility of the potential biomarker to preanalytical factors, such hemolysis or temperature of storage.

7. Conclusion
The easy accessibility to blood should not mislead scientist in their quest for new biomarkers in diseases. Its enormous complexity hides, even for the most powerful analytical technique, the needle in the haystack. In addition, it is impossible to find a unique condition reflecting perfectly the state of the sample in vivo and assuring perfectly stable and controllable pre-analytical conditions in vitro. Consequently, the discovery phase should, most of the times, be separated from the validation phase and even be done on different type of samples. The relevant clinical questions should be raised first. The practicality of the test in a real hospital world should be addressed in the design of the validation phase and particularly in the choice of sample types. In conclusion, it is anticipated that in the race for biomarkers, the cleverest will win over the more blood-flooded researcher. Another conclusion could be "errare humanum est, sed persevare diabolicum"